\documentclass[aps,twocolumn,pra,showpacs]{revtex4}
\usepackage{exscale}
\usepackage{graphicx}     
\usepackage{dcolumn}      
\usepackage{amsmath}
\usepackage{latexsym}
\usepackage{amsfonts}
\usepackage{amsmath}
\usepackage{amssymb}
\usepackage{mathrsfs}
\usepackage{amscd}
\usepackage{bm}           
\usepackage{bbm}
\usepackage{dcolumn}      
\usepackage{bm}           
\usepackage{pstricks,pst-grad,fancybox,graphics}
\usepackage{enumerate}
\usepackage{color}
\usepackage{ulem}

\newcommand{\bra}[1]{\ensuremath{\langle#1|}}

\newcommand{\ket}[1]{\ensuremath{|#1\rangle}}

\newcommand{\braket}[2]{\ensuremath{\langle #1|#2\rangle}}

\newcommand{\ketbra}[1]{\ensuremath{| #1 \rangle \langle #1 |}}

\newcommand{\Eins}{\openone}

\newcommand{\BE}{\begin{equation}}
\newcommand{\EE}{\end{equation}}
\newcommand{\be}{\begin{equation}}
\newcommand{\ee}{\end{equation}}
\newcommand{\bea}{\begin{eqnarray}}
\newcommand{\eea}{\end{eqnarray}}
\newcommand{\bean}{\begin{eqnarray*}}
\newcommand{\eean}{\end{eqnarray*}}
\newcommand{\kommentar}[1]{}

\newcommand{\mean}[1]{\ensuremath{\langle #1 \rangle}}

\newcommand{\proj}[1]{\ketbra{#1}}
\newcommand{\tr}{{\rm Tr}}

\newcommand{\bc}{\begin{center}}
\newcommand{\ec}{\end{center}}
\newcommand{\proofend}{\hfill\fbox\\\medskip }

\begin{document}

\title{Entanglement and Extreme Spin Squeezing \\
for a Fluctuating Number of Indistinguishable 
Particles}

\author{Philipp Hyllus$^{1,2}$, Luca Pezz{\'e}$^3$, Augusto Smerzi$^{2,3}$, and G{\'e}za T{\'o}th$^{1,4,5}$
}

\affiliation{
$^1$Department of Theoretical Physics, University of the Basque Country UPV/EHU, P.O. Box 644, E-48080 Bilbao, Spain\\
$^2$INO-CNR BEC Center and Dipartimento di Fisica, Universit{\`a} di Trento, I-38123 Povo, Italy\\
$^3$INO-CNR and LENS, Largo Enrico Fermi 6, I-50125 Firenze, Italy\\
$^4$IKERBASQUE, Basque Foundation for Science, E-48011 Bilbao, Spain\\
$^5$Wigner Research Center for Physics, Hungarian Academy of Sciences, P.O. Box 49, H-1525 Budapest, Hungary
}


\date{\today}

\begin{abstract}
We extend the criteria for $k$-particle entanglement from 
the spin squeezing parameter presented in [A.S. S{\o}rensen and K. M{\o}lmer,
Phys. Rev. Lett. {\bf 86}, 4431 (2001)] to systems with a fluctating number of
particles. We also discuss how other spin squeezing inequalities
can be generalized to this situation. Further, we give an operational 
meaning to the bounds for cases where the individual particles cannot be addressed. 
As a by-product, this allows us to show that in spin squeezing experiments
with cold gases the particles are typically distinguishable
in practise.
Our results justify the application of the S{\o}rensen-M{\o}lmer bounds 
in recent experiments on spin squeezing in Bose-Einstein condensates.
\end{abstract}

\maketitle

\section{Introduction}

Spin squeezing \cite{KitagawaPRA93,WinelandPRA94,MaPR11} is a central concept
in quantum metrology \cite{GiovannettiNatPhot11,BanaszeNatPhot09} and entanglement
detection \cite{GuehnePR09} in systems with a large number of 
particles. 
The most prominent spin squeezing parameter, 
defined for $N$ spin-$\frac{1}{2}$ particles or qubits, is
\cite{WinelandPRA94}
\be \label{eq:SSP}
	\xi^2=\frac{N(\Delta \hat J_{\mathbf{\perp}})^2}{\mean{\hat J_{\mathbf{n}}}^2}.
\ee
Here $\hat J_{\mathbf{n}}=\sum_{i=1}^N \hat{j}_{\mathbf{n}}^{(i)}$ is a collective spin operator
pointing along the direction $\mathbf{n}$ in the Bloch sphere, 
$\hat{j}_{\mathbf{n}}^{(i)}$ is the angular momentum operator for the 
particle $i$
and ${\mathbf{\perp}}$ is a direction perpendicular to $\mathbf{n}$.
It has been shown that a
value $\xi<1$ implies that the state of the $N$ particles is entangled \cite{SoerensenNat01}.
In addition, it allows for a phase uncertainty below the shot-noise limit, {\it i.e.,} $\Delta\theta <\frac{1}{\sqrt{N}}$ 
\cite{WinelandPRA94,PezzePRL2009,noteRamsey}, when used as input of the interferometer implementing the 
unitary transformation $e^{-i \theta \hat J_{\mathbf{m}}}$, 
where $\mathbf{m}$ is a direction perpendicular to both $\mathbf{n}$ and $\mathbf{\perp}$.

The relation between spin squeezing and entanglement has been
further extended by S{\o}rensen and M{\o}lmer in Ref.~\cite{SoerensenPRL01},
where bounds on $\xi$ have been derived 
for a partitioning of the state into groups
of at most $1\leq k < N$ particles. 
A violation of these bounds implies that 
there is at least one group of more than $k$ particles
that is fully entangled. Hence the state contains
at least $(k+1)$-particle entanglement or, according to the 
definition in Ref.~\cite{SoerensenPRL01}, an entanglement 
depth $k+1$.
The criteria were applied recent experiments 
on spin squeezing in Bose-Einstein condensates 
(BECs)~\cite{GrossNat10,RiedelNat10}. 
However, while the criteria were derived for a fixed number of
distinguishable atoms, the experiments were performed 
with a fluctuating number of bosons sharing the same trap. 
Hence the criteria have to be generalized to (i) a 
{\it non-fixed} number of (ii) {\it indistinguishable} particles.

For the case $k=1$, this has been done in
Ref.~\cite{HyllusPRL10}.
There, it has been shown that
\be \label{eq:SSP2}
\xi^2=\frac{\langle \hat N \rangle (\Delta \hat J_{\mathbf{\perp}})^2}{\mean{\hat J_{\mathbf{n}}}^2}
\ee
is a natural generalization of the spin squeezing parameter \cite{notaSpinSq}.
In particular, the condition $\xi<1$ is sufficient for sub shot-noise phase 
estimation, $\Delta \theta < \frac{1}{\sqrt{\langle \hat N \rangle}}$, 
and signals entanglement if the input state does not 
contain coherences between states with a different number of particles 
\cite{HyllusPRL10}. 
This justifies the use of the spin squeezing
parameter from Eq.~(\ref{eq:SSP2}) in experiments with cold 
\cite{AppelPNAS09,Schleier-SmithPRL10,ChenPRL11}
and ultra-cold \cite{GrossNat10,RiedelNat10} atomic gases
(for an exhaustive list see \cite{MaPR11}).
In Ref.~\cite{HyllusPRL10}, it has been argued 
that, formally, the connection between sub shot-noise 
sensitivity and entanglement holds also for 
indistinguishable particles.

In this manuscript, we extend, to the case of a fluctuating number of particles, 
the $k$-particle entanglement criteria of Ref.~\cite{SoerensenPRL01}.
We also show how the generalized spin squeezing entanglement criteria of 
Refs~\cite{TothPRL07,DuanPRL11} can be extended accordingly.
Afterwards, we use additional atomic degrees of freedom 
to extend the $k$-particle entanglement criteria to
indistinguishable particles. 
We show that in a typical 
spin squeezing experiment with cold, but not ultra-cold 
atomic gases, the particles can be treated as distinguishable 
effectively. These results apply also to other spin squeezing 
criteria \cite{WangPRA03,TothPRA04,WiesniakNJP05,KorbiczPRL05,
KorbiczPRA06,TothPRA07,TothPRL07,
TothPRA09,HePRA11,VitaglianoPRL11,DuanPRL11,GuehnePR09,MaPR11}
which are generally derived for a fixed 
number of distinguishable particles. 

The article is organized as follows. In Section \ref{sec:NonfixedN},
we discuss the generalization to a nonfixed number of particles.
In Section \ref{sec:indist}, we consider the applicability of the bounds 
for indistinguishable particles, discussing explicitly
cold atomic ensembles and BECs. 
The conclusions can be found in Section 
\ref{sec:conclusions}.

\section{Spin-squeezing bounds for a fluctuating number of particles}
\label{sec:NonfixedN}

Let us first recall the definition of $(k+1)$-particle entanglement 
and how it can be extended to the case of a fluctuating number of particles.

A pure state of $N$ particles is $k$-producible \cite{GuehneNJP05,ChenPRA05} 
if it can be written as 
\be\label{eq:psi_k}
\ket{\psi_{k-{\rm prod}}^{(N)}}=\otimes_{\alpha=1}^{M_{N}}\ket{\psi_\alpha^{(N_\alpha)}},
\ee
where $\ket{\psi_\alpha^{(N_\alpha)}}$ is a state of $N_\alpha\le k$ 
particles (such that $\sum_{\alpha=1}^{M_{N}} N_\alpha=N$).
A mixed state is $k$-producible if it can be written as a mixture 
$\rho_{k-{\rm prod}}^{(N)}=\sum_l p_l \proj{\psi_{k_l-{\rm prod}}^{(N)}}$ with $k_l\le k$ for all $l$.
A state that is $(k+1)$-producible but not
$k$-producible is referred to as $(k+1)$-particle entangled
because it contains full entanglement of at least one 
group of $k+1$ particles (with $1 \leq k < N$) \cite{SoerensenPRL01,SeevinckPRA02}.
The concept of (k+1)-particle entanglement was referred to 
as entanglement depth in Ref.~\cite{SoerensenPRL01}.

The extension of the above definition to the case of a fluctuating number of 
particles follows Ref.~\cite{HyllusPRL10}.
In this case, we define a quantum state to be $k$-producible 
{\it iff} it is $k$-producible in every fixed-$N$ subspace. 
Hence a $k$-producible quantum state without coherences 
between states of different 
$N$ can be written as
\be	\label{rho_inc}
	\rho^{\rm inc}_{k-{\rm prod}}=\sum_N Q_N \rho^{(N)}_{k-{\rm prod}},
\ee
where $\rho^{(N)}_{k-{\rm prod}}$ is a state of $N$ particles and 
$\{Q_N\}$ forms a probability distribution. 
In practice, $Q_N\to 0$ if $N$ is above some 
threshold due to energy restrictions in the lab.
For a general state $\rho$ which may contain coherences between 
different $N$, we introduce the projection 
\be\label{eq:Nproj} 
\Eins_N \rho\Eins_N=Q_N \rho^{(N)}, 
\ee
where $\Eins_N$ is the projector to the subspace of $N$ particles 
and $\rho^{(N)}$ is a state on this subspace. We may then
define a state to be $k$-producible in general if 
$\rho^{(N)}$ is $k$-producible for any $N$.

Note that there is an ongoing debate about whether
or not superpositions between states of different particle
numbers can actually be created \cite{BartlettRMP07}.
It turns out that since the angular momentum operator
$\hat J_{\mathbf{n}}=\oplus_N \hat J_{\mathbf{n}}^{(N)}$, 
for any arbitrary direction $\mathbf{n}$,
commutes with the number operator 
$\hat N = \oplus_N N\Eins_N$, such coherences do not have any effect
for entanglement detection with $\hat J_{\mathbf{n}}$ and its moments
\cite{TothPRA04}.

\subsection{Generalizing the S{\o}rensen-M{\o}lmer criteria 
to a fluctuating $N$}

Bounds on $\xi$ have been derived for states 
of $N$ spin-$j$ particles among which at most groups of $k$ 
particles are entangled \cite{SoerensenPRL01}. 
The bounds are computed with the help of the function 
\cite{noteSM}
\be\label{f_j}
	F_j(X)
	\equiv\frac{1}{j}\min_\rho(\Delta {\hat j}_{\mathbf{\perp}})^2
	\Big|_{\frac{\mean{{\hat j}_{\mathbf{n}}}}{j}=X},
\ee
where the minimization is performed over all states $\rho$ 
of a spin-$j$ particle which fulfill 
$\mean{\hat{j}_{\mathbf{n}}}/j=X$ for some $X\in[0,1]$ \cite{nota_X}.
In Eq.~(\ref{f_j}), $\hat{j}_{\mathbf{\perp}}$
and $\hat{j}_{\mathbf{n}}$ are spin operators 
for the single spin-$j$ particle.
It is then shown that for $k$-producible states,
the bound
\be\label{SSI_multi}
(\Delta\hat J_{\mathbf{\perp}})^2\ge Nj\ F_{kj}\bigg(\frac{\mean{\hat J_{\mathbf{n}}}}{Nj}\bigg).
\ee
holds, where $\hat J_{\mathbf{n}}=\sum_{i=1}^N\hat j_{\mathbf{n}}^{(i)}$
and in analogy for $\hat J_{\mathbf{\perp}}$, as introduced above.
Hence if the measured values of $\mean{\hat J_{\mathbf{n}}}$ and 
$(\Delta\hat J_{\mathbf{\perp}})^2$ violate Eq.~(\ref{SSI_multi}), 
then the state is at least $(k+1)$-particle entangled.
For $j=\frac{1}{2}$, the state allows for a smaller uncertainty in an
interferometric protocol than any $k$-producible
state.

Before generalizing these bounds to states with a nonfixed $N$,
we remark that a different method, based on the Quantum Fisher information, 
to detect $(k+1)$-particle entanglement for a state of a fixed number of particles,
has been recently introduced in Ref.~\cite{HyllusPRA2012}.

\noindent
{\bf Observation 1}. 
{\it 
For $k$-producible states of spin-$j$ particles with a fluctuating total number,
and with given average values $\mean{\hat N}$ and 
$\mean{\hat J_{\mathbf{n}}}$, the inequality 
\be \label{eq:Obs1}
	(\Delta\hat J_{\mathbf{\perp}})^2\ge \mean{\hat N}j\ F_{kj}
	\bigg(\frac{\mean{\hat J_{\mathbf{n}}}}{\mean{\hat N}j}\bigg)
\ee
holds, irrespectively of whether or not coherences between different numbers
of particles are present in the state. }

The proof is given in Appendix A. Note that Eq.~(\ref{eq:Obs1}) 
reduces to Eq.~(\ref{SSI_multi}) for a fixed number of particles.
Also for a fixed $N$, Observation 1 
extends the seminal result of Ref.~\cite{SoerensenPRL01} in two ways.
Firstly, in the proof, a step is carried out 
[below Eq.~(\ref{eq:Obs1proof1})] 
which was not discussed explicitly in the original proof.
Further, Observation 1 does not require $\frac{N}{k}$ to be an integer 
as in the original criterion. In order to apply it for non-fixed $N$, 
simply $N$ has to replaced by $\mean{\hat N}$, as in the usual spin squeezing 
criterion \cite{HyllusPRL10}.

\subsection{Generalizing other spin squeezing inequalities to a fluctuating $N$}
\label{sec:OurSSI4nonfixedN}

We now consider
other spin squeezing inequalities for entanglement detection \cite{WangPRA03,TothPRA04,WiesniakNJP05,KorbiczPRL05,KorbiczPRA06,TothPRA07,
TothPRL07,TothPRA09,HePRA11,VitaglianoPRL11,DuanPRL11,GuehnePR09,MaPR11},
which have been derived for a fixed number of particles.
Most of them can be generalized to the case of a fluctuating number of particles by 
directly using the inequality
\bea
	(\Delta \hat J_{\mathbf{n}})^2&=& \sum_N Q_{N}\mean{(\hat J_{\mathbf{n}}^{(N)})^2} - 
	\big(\sum_N Q_{N}\mean{\hat J_{\mathbf{n}}^{(N)}}\big)^2 \nonumber\\
	&\ge &  \sum_N Q_{N} (\Delta \hat J_{\mathbf{n}}^{(N)})^2, \label{eq:VarConcavity}
\eea
which can be derived using the Cauchy-Schwarz inequality, 
for any arbitrary direction $\mathbf{n}$. It also 
follows from the concavity of the variance. 
Further, note that for any operator $\hat{O}=\oplus_{N=0}^\infty \hat{O}^{(N)}$, which commutes
with $\hat N$,
\be\label{eq:Ablockdiag}
 \mean{\hat O^l}=\tr[\rho \hat O^l]=\sum_N Q_N \tr[\rho^{(N)} (\hat O^{(N)})^l]
\ee
holds for any power $l$. 
All angular momentum operators $\hat J_{\mathbf{n}}$ are of this form. 
Therefore, coherences between states of different $N$ in $\rho$ 
do not play any role in entanglement detection with
any kind of spin squeezing criteria, as mentioned above.

As an example, we perform the generalization 
for the complete set of inequalities 
from Ref.~\cite{TothPRL07}
and for the criteria detecting $k$-particle entanglement
from Ref.~\cite{DuanPRL11}.
All these criteria have been derived for $N$ 
particles with spin $j=\frac{1}{2}$.
The set of criteria of Ref.~\cite{TothPRL07} is
\bea
	\mean{\hat J_{\mathbf{x}}^2}+\mean{\hat J_{\mathbf{y}}^2}+\mean{\hat J_{\mathbf{z}}^2}&\le&N(N+2)/4
	\label{eq:gSSI1}\\
	(\Delta \hat J_{\mathbf{x}})^2+(\Delta \hat J_{\mathbf{y}})^2+(\Delta \hat J_{\mathbf{z}})^2&\ge& N/2
	\label{eq:gSSI2}\\
	\mean{\hat J_{\mathbf{i}}^2}+\mean{\hat J_{\mathbf{j}}^2}-N/2&\le&(N-1)(\Delta \hat J_{\mathbf{l}})^2
	\label{eq:gSSI3}\\
	(N-1)[(\Delta \hat J_{\mathbf{i}})^2+(\Delta \hat J_{\mathbf{j}})^2]&\ge&\mean{\hat J_{\mathbf{l}}^2}+N(N-2)/4,\nonumber\\
  &&	\label{eq:gSSI4}
\eea
where ${\mathbf{i}},{\mathbf{j}},{\mathbf{l}}$ take all possible permutations of 
${\mathbf{x}},{\mathbf{y}},{\mathbf{z}}$. 
This set 
is complete in the sense that it detects all entangled states
which can be detected based on the knowledge of $\mean{\hat J_i^2}$
and $(\Delta \hat J_{\mathbf{i}})^2$ for ${\mathbf{i}}={\mathbf{x}},{\mathbf{y}},{\mathbf{z}}$ \cite{TothPRL07}.

Due to linearity, inequality~(\ref{eq:gSSI1}), which is valid for
all quantum states, directly generalizes to
\be
	\mean{\hat J_{\mathbf{x}}^2}+\mean{\hat J_{\mathbf{y}}^2}+\mean{\hat J_{\mathbf{z}}^2}
	\le[\mean{\hat N^2}+2\mean{\hat N}]/4.
\ee
Inequality~(\ref{eq:gSSI2}) can be generalized using Eq.~(\ref{eq:VarConcavity})
to 
\be
	(\Delta \hat J_{\mathbf{x}})^2+(\Delta \hat J_{\mathbf{y}})^2+(\Delta \hat J_{\mathbf{z}})^2\ge \mean{\hat N}/2.
\ee
In analogy, the inequalities~(\ref{eq:gSSI3},\ref{eq:gSSI4}) 
can be generalized by applying Eq.~(\ref{eq:VarConcavity}) to the 
variances. The result can be written as 
\bea
	(\Delta\hat J_{\mathbf{l}})^2 &\ge&\mean{(\hat N-1)^{-1}\hat J_{\mathbf{i}}^2}\label{eq:gSSI_N1}\\
	&&+\mean{(\hat N-1)^{-1}\hat J_{\mathbf{j}}^2}-\mean{(\hat N-1)^{-1}\hat N}/2\nonumber\\
	(\Delta\hat J_{\mathbf{i}})^2+(\Delta\hat J_{\mathbf{j}})^2 &\ge& \mean{(\hat N-1)^{-1}
	\hat J_{\mathbf{l}}^2}\label{eq:gSSI_N2}\\
	&&+\mean{(\hat N-1)^{-1}\hat N(\hat N-2)}/4.\nonumber
\eea
Here it is assumed that $Q_0=Q_1=0$. This should not pose a problem
because the spin squeezing criteria are developed for a large number
of particles. A conceptual change in the generalized criteria from
Eqs~(\ref{eq:gSSI_N1}) and (\ref{eq:gSSI_N2}) is that instead of the
expectation values $\mean{\hat J_{\mathbf{i}}^2}$, terms such as $\mean{(\hat N-1)^{-1}\hat J_{\mathbf{i}}^2}$
appear. This implies that the number of particles has to be measured
in each shot, which might complicate the application in some experiments.
In the same way, the set of inequalities for $N$ spin-$j$ particles from
Ref.~\cite{VitaglianoPRL11} can be generalized to a non-fixed $N$. 

Note that alternatively, the criteria could be tested for a fixed number of particles $N$. 
In this case, one could collect separate statistics for each $N$. If the number fluctuates 
strongly, it would be very difficult to collect enough statistics for a given fixed $N$, 
while it is still possible to have enough statistics for the generalized criteria.

We finally remark that the bound 
\be
	(\Delta\hat J_{\mathbf{z}})^2 \ge \frac{1}{k+2}\bigg[\frac{\mean{\hat J_{\mathbf{x}}^2}}{N}
	+\frac{\mean{\hat J_{\mathbf{y}}^2}}{N}\bigg]-\frac{1}{4}
\ee
for $k$-producible states from Ref.~\cite{DuanPRL11}
can be generalized to 
\be\label{eq:Duan_gen}
	(\Delta\hat J_{\mathbf{z}})^2 \ge \frac{1}{k+2}\big[\mean{\hat N^{-1}\hat J_{\mathbf{x}}^2}
	+\mean{\hat N^{-1}\hat J_{\mathbf{y}}^2}\big]-\frac{1}{4}.
\ee
This bound is optimal for the symmetric twin-Fock states with 
$N/2$ particles in each of the two modes of an interferometer which 
promises a phase uncertainty close to the ultimate
Heisenberg limit, $\Delta \theta = \frac{1}{N}$ \cite{HollandPRL93}. Recently, such states have 
been prepared experimentally with ultra-cold atomic gases 
\cite{BookjansPRL11,BueckerNatPhys11,LueckeSci11,GrossNat11}. Since the number
of atoms fluctuates in these experiments, Eq.~(\ref{eq:Duan_gen})
could be used to bound $k$, while Eq.~(\ref{eq:Obs1}) 
from Observation 1 is generally not useful in this situation since 
$\mean{\hat J_{\mathbf{z}}}=0$ for these states. However, the same problem 
concerning the indistinguishability
of the particles occurs also here. This problem will be discussed
in the next Section.

\section{Spin-squeezing bounds for indistinguishable particles}
\label{sec:indist}

The bounds (\ref{SSI_multi}) and (\ref{eq:Obs1}) presented above 
have been derived for distinguishable particles. 
This corresponds to the usual situation employed in 
quantum information theory with, for instance, trapped ions. In this case 
the particles are assumed to sit at remote locations
and operations are only performed on the internal degrees of freedom, locally at
each trap. The particles can be treated as distinguishable, labelled by the 
trap number, and the (anti-)symmetrization can be dropped \cite{PeresBook}.

However, Eq.~(\ref{SSI_multi})
has been recently applied to discuss spin squeezing
experiments with Bose-Einstein condensates \cite{GrossNat10,RiedelNat10}.
In this situation, all the particles (bosons)
share the same trap state. Their collective internal state has to 
be fully symmetric with respect to the interchange of any two particles in first quantization. 
For indistinguishable bosons, the (symmetric) fully separable states
have the form $\ket{\phi}^{\otimes N}$.
The spin squeezing condition $\xi<1$, see Eqs. (\ref{eq:SSP}) and (\ref{eq:SSP2}),
still holds and signals entanglement in the sense that the state of the indistinguishable 
bosons cannot be written as $\ket{\phi}^{\otimes N}$.
The relation between shot-noise limit and separable states holds formally as well 
\cite{HyllusPRA10,HyllusPRL10,nota_modes,nota_indist}.

In contrast, a symmetric state of $N$ particles can be either fully separable or
fully entangled, but no symmetric states that 
are $k$-particle entangled as in Eq.~(\ref{eq:psi_k}) exist for $1<k<N$
\cite{EckertAoP02,IchikawaPRA08,WeiPRA10}. 
Hence the classification introduced above for distinguishable particles is 
not directly applicable to recent experiments with BECs, where the individual 
particles are not addressable.
The same problem would occur if the criteria for $k$-particle
entanglement proposed in \cite{DuanPRL11} and generalized in Eq.~(\ref{eq:Duan_gen})
were applied to the twin-Fock states produced recently 
with ultra-cold atomic gases \cite{BookjansPRL11,BueckerNatPhys11,LueckeSci11,GrossNat11}.

\subsection{Entanglement and spin squeezing due to symmetrization} 

First, let us notice that
the collective spin operators $\hat J_{\mathbf{n}}$, which appears
in the definition of the spin squeezing parameter $\xi$ [Eqs.~(\ref{eq:SSP})
and (\ref{eq:SSP2})], are 
permutationally invariant, {\it i.e.}, 
$\hat P_\pi^\dagger \hat J_{\mathbf{n}}\hat P_\pi=\hat J_{\mathbf{n}}$
for any of the $N!$ permutations $\pi$ of the $N$ particles (represented by $\hat P_\pi$).
Therefore, $\tr[\rho\hat J_{\mathbf{n}}] = \tr[\rho_{\rm PI}\hat J_{\mathbf{n}}]$,
where $\rho_{\rm PI}=\frac{1}{N!}\sum_\pi \hat P_\pi^\dagger \rho \hat P_\pi$
is permutationally invariant. 
One may think that, because of this property of the collective spin operators, 
the spin squeezing bounds for non-symmetric and corresponding symmetrized states 
should remain the same. 
However, a state of $N$ bosons needs not only to be permutationally 
invariant, but symmetric with respect to the interchange
of any two particles, {\it i.e.}, it has to be possible to write it 
as a mixture of symmetric pure states fulfilling 
$\hat P_\pi\ket{\psi_{\rm S}}=\ket{\psi_{\rm S}}$ for any permutation
$\pi$. This is a much stronger requirement \cite{TothPRL09}. 

Consider, for example, the 
permutationally invariant state of $N=2$ particles
$\varrho_{\rm PI}=[\proj{0}\otimes\proj{1}+\proj{1}\otimes\proj{0}]/2$.
Here $\ket{0}$ and $\ket{1}$ are eigenstates of the Pauli matrix $\hat\sigma_z$
with eigenvalue +1 and -1, respectively. 
This can be rewritten as 
$\varrho_{\rm PI}=[\proj{\psi^+}+\proj{\psi^-}]/2$, where
$\ket{\psi^\pm}=(\ket{01}\pm\ket{10})/\sqrt{2}$. Hence
$\varrho_{\rm PI}$ does not live on the symmetric subspace
because it has an antisymmetric component $\ket{\psi^-}$.
Since the state is separable, $\xi\ge 1$ for 
any combination of the directions ${\mathbf{n}}$ and 
${\mathbf{\perp}}$. Hence it does not allow for sub shot-noise
phase estimation. Projecting $\varrho_{\rm PI}$ onto the
symmetric subspace leads to 
$\varrho_{\rm PI}\to \ket{\psi^+}$. This state is known as 
a twin-Fock state of $N=2$ particles \cite{HollandPRL93}.
It is entangled \cite{HyllusPRA10} and allows for sub shot-noise phase estimation
\cite{HollandPRL93} even though it is not spin-squeezed
because 
$\bra{\psi^+} \hat J_{\mathbf{n}} \ket{\psi^+}=0$ for any ${\mathbf{n}}$. 
In Appendix B,
we consider an additional example where a separable state is transformed
into an entangled spin squeezed state by symmetrization.

This shows that symmetrization does not preserve neither 
entanglement nor spin squeezing. 
In general, symmetrization does not preserve the 
$k$-producibility class of a state of $N$ particles. A
$k$-producible state will generally be $N$-particle entangled
after the symmetrization, and the bounds for a given 
$k$ do not apply anymore.

\subsection{
Generalizing the S{\o}rensen-M{\o}lmer criteria for indistinguishable particles}
\label{sec:Operational_Interpretation}

We assume that 
the collective spin transformations and measurements are 
performed on two energy levels of each atom, which we refer to as 
the internal degrees of freedom.
The extension of the bounds (\ref{SSI_multi}) and (\ref{eq:Obs1}) to 
indistinguishable particles is based on the inclusion of the atomic
external degrees of freedom such as the spatial trap states. 
We thus consider operations 
of the form $\hat A_{\rm in}\otimes\Eins_{\rm ex}$, where $\hat A_{\rm in}$ 
acts on the internal degrees of freedom and $\Eins_{\rm ex}$ is the identity acting
on the external degrees of freedom. 
The operator $\hat A_{\rm in}\otimes\Eins_{\rm ex}$ must be permutationally 
invariant because we consider indistinguishable particles \cite{PeresBook}.  
As mentioned above, this is the case for the collective 
spin operators $\hat A_{\rm in} = \hat J_{\mathbf{n}}$.

The basic idea is that particles can be distinguished 
here by their external state. 
Therefore, the state needs to be symmetrized only with respect 
to all particles in the {\it same} external state, but not with respect
to particles in {\it different} external states.
This is true even though the operations introduced 
above do not resolve the external states \cite{FeynmanLectures3}. 

Let us illustrate this with an example. We consider $N=2$ particles,
labeled as $1$ and $2$,
in two different external states, labeled as $a$ and $b$ ($\braket{a}{b}=0$). 
Following Ref.~\cite{PeresBook}, a general pure symmetric state can be written as 
\be\label{eq:general_state}
	\ket{\psi}=\frac{1}{\sqrt{2}}\big(\ket{\psi_{12}}_{\rm in}\otimes\ket{a_1 b_2}_{\rm ex}+\ket{\psi_{21}}_{\rm in}\otimes\ket{b_1 a_2}_{\rm ex}\big),
\ee
where $\ket{\psi_{12}}_{\rm in}$ is a general (not necessarily symmetric) internal state of the two particles, 
$\ket{\psi_{21}}_{\rm in}=\hat{P}_{\rm in}\ket{\psi_{12}}_{\rm in}$, $\hat{P}_{\rm in}$
permutes the particles, and $\ket{a_i b_j}_{\rm ex}$ is the external (e.g. spatial) wave function
($i,j=1,2$, $i \neq j$).  
The mean value of the operator $\hat A_{\rm in}\otimes\Eins_{\rm ex}$ is
\bea
	\bra{\psi}\hat A_{\rm in}\otimes\Eins_{\rm ex}\ket{\psi}=
	\frac{\bra{\psi_{12}}\hat A_{\rm in}\ket{\psi_{12}}
	             +\bra{\psi_{21}}\hat A_{\rm in}\ket{\psi_{21}}}{2},
\eea
where the two terms in the sum are equal since 
$\hat A_{\rm in}$ is permutationally invariant:
\be
\bra{\psi_{21}} \hat A_{\rm in} \ket{\psi_{21}} = 
\bra{\psi_{12}} \hat{P}_{\rm in}^\dag \hat A_{\rm in} \hat{P}_{\rm in} \ket{\psi_{12}} 
= \bra{\psi_{12}} \hat A_{\rm in} \ket{\psi_{12}}.
\ee
We dropped the label ``in'' of $\ket{\psi_{12}}_{\rm in}$
for simplicity.
The above equations show that $\ket{\psi_{12}}$ is sufficient to describe 
the state of the two particles.
In particular, non-symmetric states $\ket{\psi_{12}}$ are allowed and
the two particles can be formally treated
as distinguishable. 

The generalization to a system of $N_\gamma$ particles in the external
level $\gamma$ (such that $\sum_\gamma N_\gamma=N$)
can be formulated as follows.

\noindent
{\bf Observation 3}. {\it The expectation value }
$\mean{\hat A_{\rm in}\otimes \Eins_{\rm ex}}$ 
{\it for 
any permutationally invariant operator $\hat A_{\rm in}$
with respect to a fully symmetric state $\ket{\psi}$ with
$N_\gamma$ bosons in the external state $\gamma$ is equal
to the expectation value $\mean{\hat A_{\rm in}}$ 
computed with respect to the corresponding internal state $\ket{\psi}_{\rm in}$,
which is symmetrized only with respect to
the particles sharing the same state $\gamma$, for all $\gamma$.
}

\begin{figure}[t!]
\begin{center}
\includegraphics[clip,scale=0.6]{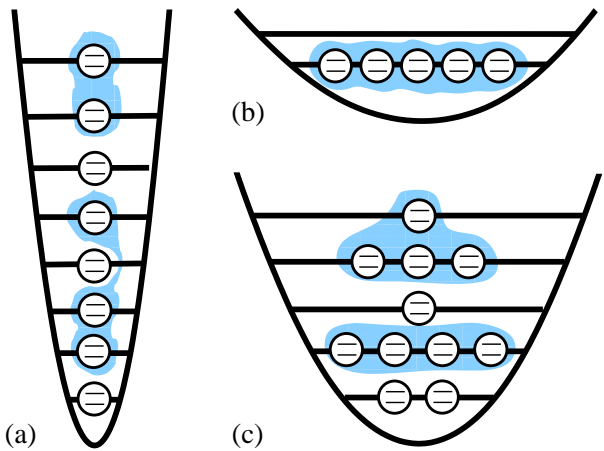}
\end{center}
\caption{
Illustration of Observation 3 with particles of 
$d_{\rm in}=2$ internal states.
The light-gray structure (color online) indicates
entanglement between particles.
(a) Fully distinguishable particles ($d_{\rm ex}=N$).
Depicted is a $3$-particle entangled state of $N=8$ particles.
Entanglement is present
between the particles with energy level $\gamma$ equal to
2,3, and 5 (counting upwards from the lowest level with $\gamma=1$), 
and between particles 7 and 8.
(b) Fully indistinguishable particles ($d_{\rm ex}=1$).
Depicted is a state which does not factorize.
(c) Mixed situation ($1<d_{\rm ex}<N$).
Depicted is a $4$-particle entangled state of $N=11$ particles in
$d_{\rm ex}=5$ external levels. 
There are two groups of 4 fully entangled particles: 
in the external level $\gamma=2$, where the 
4 particles are indistinguishable and 
in a non-factorisable state; and in levels 4 and 5. In the latter case,
the group of particles in level $\gamma=4$ is distinguishable from
the particle in level $\gamma=5$. 
The two particles in the lowest level $\gamma=1$ are in a symmetric 
separable state of the form $\ket{\phi}\otimes\ket{\phi}$.
}  \label{fig:k_particle_entangled_states}
\end{figure}

This Observation is formulated more precisely in
Appendix C below, where also the relationship of $\ket{\psi}$ and 
$\ket{\psi}_{\rm in}$ is explained in detail. 

Figure~\ref{fig:k_particle_entangled_states} illustrates several examples of 
$N$ particles in $d_{\rm in}=2$ internal and $d_{\rm ex}$ external modes.
The usual situation employed in quantum information theory, 
where all particles are distinguishable ($d_{\rm ex}=N$), is shown in 
Fig.~\ref{fig:k_particle_entangled_states}(a).
For what concerns our discussion, this is formally equivalent to an 
array of separated wells, as in ions traps.
The opposite situation of all particles occupying the same level
($d_{\rm ex}=1$), is shown in 
Fig.~\ref{fig:k_particle_entangled_states}(b).
For indistinguishable particles, only two possibilities are allowed in this case: 
either all particles are in a separable ({\it i.e.}, product $\ket{\phi}^{\otimes N}$)
state, or all particles are entangled, due to the symmetrization 
\cite{EckertAoP02,IchikawaPRA08,WeiPRA10}.
As mentioned above, the $k$-particle entanglement criterion discussed in Sec.~\ref{sec:NonfixedN}
does not apply in this case.
The interesting intermediate situation is shown in 
Fig.~\ref{fig:k_particle_entangled_states}(c).
In this case several particles may occupy the same external state. 
As noticed in Observation 3, the symmetrization is necessary only for particles 
that share the same external level $\gamma$.
In this case, the $N_\gamma$ particles may be only found in a fully entangled 
or fully separable state. 
However, entanglement is also possible between particles occupying different levels.

We can now extend the S{\o}rensen-M{\o}lmer bounds.
A state can be considered as (effectively) $k$-producible if
\be \label{eq:psi_k_Obs2}
\ket{\psi}_{\rm in}=\otimes_{\alpha=1}^M \ket{\psi_\alpha}.
\ee
where $\ket{\psi_\alpha}$ is state of $N_\alpha \le k$ particles
($\sum_{\alpha=1}^M N_\alpha =N$) for all $\alpha$.
The particles in the state $\ket{\psi_\alpha}$
can occupy a single external state $\gamma$ (in which case $N_\alpha = N_\gamma$ 
and $\ket{\psi_\alpha}=\ket{\psi_\gamma}$ is symmetric) or
different external states $\gamma\in I_\alpha$
(in which case $N_\alpha =\sum_{\gamma\in I_\alpha} N_\gamma$ and
$\ket{\psi_\alpha}$ is not necessarily symmetric).
As an example,
the state schematically shown in 
Fig.~\ref{fig:k_particle_entangled_states}(c)
is 4-particle entangled.

With this notion, the S{\o}rensen-M{\o}lmer criteria can 
be applied in systems of indistinguishable particles as follows.

\noindent
{\bf Observation 4}. {\it 
(i) For particles of spin $j$, if the spin squeezing parameter 
violates Eq.~(\ref{SSI_multi}) for a 
given $k$, then the input state cannot be written
as a mixture of effectively $k$-producible states of 
Eq.~(\ref{eq:psi_k_Obs2}).
(ii) For particles of spin $\frac{1}{2}$, if the spin squeezing parameter 
violates Eq.~(\ref{SSI_multi}) for a 
given $k$, then the input state allows for a smaller
phase uncertainty than the smallest one achievable with 
a mixture of effectively $k$-producible states of Eq.~(\ref{eq:psi_k_Obs2}).
}

In both cases, effective $(k+1)$-particle 
entanglement is proven by a violation
of the criteria.
These notions directly generalize to systems of a fluctuating 
number of particles as in Section~\ref{sec:NonfixedN}.

\subsection{Cold atoms}

Observation 3 is also useful in the context of entanglement detection 
with generalized spin squeezing inequalities (SSI) \cite{GuehnePR09,MaPR11} 
in cold atomic clouds.  
Usually the atomic ensembles are not ultracold, and can be assumed to be in 
a thermal state externally. We estimate the population of 
the trap levels using the statistics of an ideal Bose gas
taking the parameters from a typical experiment with a 
cigar-shaped configuration \cite{KubasikPRA09,Mitchell_pc}: 
$\omega_z=2$, $\omega_{\mathbf{\perp}}=1000$ 
(trap frequencies in units of $2\pi$s$^{-1}$), 
$T=30\mu$K, $N=5\cdot10^5$. The chemical potential $\mu$
is defined implicitly by the relation $N=\sum_k \mean{n_k}$, 
where \cite{SchwablBook}
\be \label{eq:Bose}
\mean{n_k}=\sum_{n} \Big[\frac{e^{-\beta n (E_k-\mu)}}{\sum_{n'} e^{-\beta n' (E_k-\mu)}}\Big]  n = \frac{1}{e^{\beta(E_k-\mu)}-1}
\ee
is the average population of the level $k$, $\beta=1/k_B T$ with the 
the Boltzmann constant $k_B$, and $E_k$ is the energy of level $k$. 
We approximate this by the energy levels of a 3d harmonic oscillator
with the given trap frequencies.

The largest average population is obtained for the ground state 
with energy $E_0=\hbar(\omega_z/2+\omega_{\mathbf{\perp}})$. 
We obtain $\beta\mu \approx -12.41$, which leads to 
$\mean{n_0}\approx 4.1\cdot 10^{-6}$. Further,
the probability of having $n_k$ particles in the level $k$,
given by the expression in square brackets in Eq.~(\ref{eq:Bose}),
decreases exponentially with $n_k$. The ratio
$p_{n_k+1}/p_{n_k}=e^{-\beta(E_k-\mu)}$ is the largest 
for $k=0$, where it is equal to $4.1\cdot 10^{-6}$. 
Therefore, it can be assumed that at most one particle occupies
each level. Since only internal quantities are used in the
generalized SSI, one can therefore treat 
the particles as distinguishable by using Obs. 3.
This situation corresponds to the one 
depicted in 
Fig.~\ref{fig:k_particle_entangled_states}(a).

\subsection{Bose-Einstein condensates}
\label{sec:Arg1}

In Bose-Einstein condensates particles share the same external state, 
as schematically illustrated in 
Fig.~\ref{fig:k_particle_entangled_states}(b).
In this case, it is not possible to directly apply the bounds (\ref{SSI_multi}) 
and (\ref{eq:Obs1}): particle entanglement 
is either absent or maximal, due to symmetrization \cite{EckertAoP02,IchikawaPRA08,WeiPRA10}.

However, we recall that Eqs.~(\ref{SSI_multi}) and (\ref{eq:Obs1}) are sufficient 
conditions for entanglement and, as mentioned in the introduction, 
for spin $j=\frac{1}{2}$ particles,
they are strongly 
related to the usefulness of the entangled state for parameter estimation.
Therefore, a measurement of the spin squeezing bounds in BECs
might still be consistent with $k<N$, even if, by some other means, it is
possible to show that {\it all} particles share the same
external state, as in Ref.~\cite{GrossNat10}. 
The outcome $k<N$ should be interpreted either as due to noise or by 
saying that all particles are entangled but the state is only partially useful for parameter estimation.
It is known, indeed, that there are symmetric states which are fully
$N$-particle entangled, but that are not spin squeezed and do not allow for 
sub shot-noise phase estimation \cite{HyllusPRA10}.

Finally, one might think that, by making the BEC cloud very dilute, it is 
possible to effectively distinguish the particles and
thus use the spin squeezing bounds (\ref{SSI_multi}) and (\ref{eq:Obs1}).
In Appendix D we show that this is not the case.

\section{Conclusions} 
\label{sec:conclusions}

The spin squeezing criteria introduced by 
S{\o}rensen and M{\o}lmer for $N$ distinguishable particles in Ref.~\cite{SoerensenPRL01}
are a powerful and experimentally feasible method to detect $k$-particle entanglement, 
also referred to as entanglement depth $k$.
However, most of the spin squeezing experiments are performed with a fluctuating number 
of particles and, as in the case of BEC, these particles are indistinguishable.
To fill this gap between theory and experiment, 
we have extended, in the first part of this article, 
the S{\o}rensen and M{\o}lmer criteria 
to systems with a fluctuating number of particles. 
We have also shown how other spin squeezing inequalities \cite{TothPRL07,DuanPRL11}
can be generalized to this situation.
In the second part of the paper, we discussed the conceptual 
problems that occur when the individual particles are indistinguishable. 
In this case, effective $k$-particle entanglement can be defined only by 
making use of additional 
degrees of freedom of the atoms.
The spin squeezing bounds of Ref.~\cite{SoerensenPRL01}
can then be interpreted as conditions of such effective $k$-particle entanglement.
Our results make it possible to apply
the bounds of Ref.~\cite{SoerensenPRL01} 
in spin squeezing experiments with cold atoms 
and Bose-Einstein condensates.

\section*{Acknowledgements}
P.H. thanks M.W. Mitchell, K. M{\o}lmer, M. Oberthaler, F. Piazza, and J. Siewert for 
inspiring discussions. 
P.H. and G.T. acknowledge financial support of the ERC Starting Grant GEDENTQOPT
and the EU grant CHIST-ERA QUASAR.
G.T. thanks the MICINN (Project No.  FIS2009-12773-C02-02),  the  Basque  Government
(Project No. IT4720-10), and the National Research Fund
of Hungary OTKA (Contract No. K83858).
L.P. acknowledges financial
support by MIUR through FIRB Project No. RBFR08H058.

\section*{APPENDIX}

\subsection*{A. Proof of Observation 1} 

The proof follows the lines of 
the proof of Eq.~(\ref{SSI_multi}) for fixed $N$ 
\cite{SoerensenPRL01} using methods developed for non-fixed $N$ in Ref.~\cite{HyllusPRL10}. 
We want to compute a lower bound on the variance of $\hat J_{\mathbf{\perp}}$ for all $k$-producible states. 
Since the variance is concave in the state, 
its minimum value is reached by pure states of the form   
$\ket{\psi_{k-{\rm prod}}}=\sum_N\sqrt{Q_N}\ket{\psi_{k-{\rm prod}}^{(N)}}$ 
\cite{nota_optim}, where $\sqrt{Q_N}$ are real numbers with
$\sum_N Q_N =1$ and 
$\ket{\psi_{k-{\rm prod}}^{(N)}}=\bigotimes_{\alpha=1}^{M_N} \ket{\psi_\alpha^{(N_\alpha)}}$
is a $k$-producible state of $N$ particles [cf. Eq.~(\ref{eq:psi_k})].
Using Eq.~(\ref{eq:VarConcavity}) we can write
$(\Delta\hat J_{\mathbf{\perp}})^2\ge\sum_N Q_N (\Delta\hat J_{\mathbf{\perp}}^{(N)})^2$.
In addition, we note that, due to the product structure of the states 
$\ket{\psi_{k-{\rm prod}}^{(N)}}$ and
since $\hat J_{\mathbf{\perp}}^{(N)}$ is the sum of operators acting on fixed $N_\alpha$ subspaces, 
$\hat J_{\mathbf{\perp}}^{(N)} = \sum_{\alpha=1}^{M_N} \hat J_{\mathbf{\perp}}^{(N_\alpha)}$,
we have $(\Delta\hat J_{\mathbf{\perp}}^{(N)})^2 = \sum_{\alpha=1}^{M_N} \big(\Delta\hat J_{\mathbf{\perp}}^{(N_\alpha)} \big)^2$.
Therefore, the variance of $\hat J_{\mathbf{\perp}}$ for 
$k$-producible states is bounded by
\be\label{eq:Obs1proof1}
	(\Delta\hat J_{\mathbf{\perp}})^2 \ge 	
	\sum_N Q_{N} \sum_{\alpha=1}^{M_N} \big(\Delta\hat J_{\mathbf{\perp}}^{(N_\alpha)} \big)^2, 
\ee
where the operator on the right hand side acts on the $N_\alpha$
particles in the state $\ket{\psi_\alpha^{(N_\alpha)}}$.
Note that we did not
attach an index $N$ to $N_\alpha$ in order to simplify the notation.

Now we have to find the minimal bound for the variances
$(\Delta\hat J_{\mathbf{\perp}}^{(N_\alpha)})^2$ for every $N$, given 
the mean value $\mean{\hat J_{\mathbf{n}}^{(N_\alpha)}}$. 
If we consider $N_\alpha$ spin-$j$ particles, 
the total spin $j_\alpha$ can range from $0$ (if $N_\alpha$ 
is even) or $1/2$ (if $N_\alpha$ is odd) up to $N_\alpha \, j$.
We show that for any 
$\mean{\hat J_{\mathbf{n}}^{(N_\alpha)}}$, the smallest bound is reached 
by choosing the largest total spin possible
by using that $(\Delta\hat J_{\mathbf{\perp}}^{(N_\alpha)})^2\ge j_\alpha F_{j_\alpha}(\mean{\hat J_{\mathbf{n}}^{(N_\alpha)}}/j_\alpha)$ for states with a fixed spin $j_\alpha$,
cf. Eq.~(\ref{f_j}). 
The ingredients needed for this, 
which have been proven in Ref.~\cite{SoerensenPRL01}, 
are: 
(i) the functions $F_{j_{\alpha}}$ are convex, {\it i.e.},
$F_{j_{\alpha}}(aX+bY)\le aF_{j_{\alpha}}(X)+bF_{j_{\alpha}}(Y)$ 
for all $j_{\alpha}$ and $a,b\ge 0$ with $a+b=1$,
(ii) $F_{j_{\alpha}}(0)=0$ for all $j_{\alpha}$, 
and (iii) that $F_{j_{\alpha}}(X)\le F_{j_{\alpha}'}(X)$
if $j_{\alpha} \ge j_{\alpha}'$.
By using the inequality (i) with $a=\frac{j_{\alpha}'}{j_{\alpha}}$, $Y=0$ and $b=1-a$,
and the property (ii), we have 
$F_{j_{\alpha}}(\frac{j_{\alpha}'}{j_{\alpha}}X)\le j_{\alpha}' F_{j_{\alpha}}(X) / j_{\alpha}$.
Multiplying by $j_{\alpha}$ both terms and using (iii), we arrive at
$j_{\alpha} F_{j_{\alpha}}(\frac{j_{\alpha}'}{j_{\alpha}}X )\le j_{\alpha}' F_{j_{\alpha}'}(X)$ 
if $j_{\alpha}'\le j_{\alpha}$.
Finally, taking $X=\frac{\mean{\hat J_{\mathbf{n}}^{(N_\alpha)}}}{j_{\alpha}'}$, we have 
\be\label{eq:larger_j_is_better}
j_{\alpha} F_{j_{\alpha}}\Big(\frac{\mean{\hat J_{\mathbf{n}}^{(N_\alpha)}}}{j_{\alpha}}\Big)\le 
j_{\alpha}' F_{j_{\alpha}'}\Big(\frac{\mean{\hat J_{\mathbf{n}}^{(N_\alpha)}}}{j_{\alpha}'}\Big)
\ee 
if $j_{\alpha}\ge j_{\alpha}'$. 
Let us now consider a superposition
$\ket{\psi}=c_{j_{\alpha}}\ket{\psi_{j_{\alpha}}}+c_{{j'_{\alpha}}}\ket{\psi_{{j'_{\alpha}}}}$
of states with a different fixed spin ${j_{\alpha}}\ge {j'_{\alpha}}$.
Since the spin operator $\hat J_{\mathbf{\perp}}$ does
not couple the states of different total spin ${j_{\alpha}}$ 
and ${j'_{\alpha}}$, its variance with respect to $\ket{\psi}$
is equal to the variance with respect to the mixture
$\rho=|c_{j_{\alpha}}|^2\proj{\psi_{j_{\alpha}}}+|c_{{j'_{\alpha}}}|^2\proj{\psi_{{j'_{\alpha}}}}$.
Using the concavity of the variance,
we obtain that 
\bean
(\Delta\hat J_{\mathbf{\perp}})^2_{\ket{\psi}}&\ge&
|c_{j_{\alpha}}|^2(\Delta\hat J_{\mathbf{\perp}})^2_{\ket{\psi_{j_{\alpha}}}}+|c_{{j'_{\alpha}}}|^2(\Delta\hat J_{\mathbf{\perp}})^2_{\ket{\psi_{{j'_{\alpha}}}}}\\
&\ge&|c_{j_{\alpha}}|^2 F_{j_{\alpha}}(\mean{\hat J_{\mathbf{n}}}_{\ket{\psi}}/{j_{\alpha}})+|c_{{j'_{\alpha}}}|^2 
F_{{j'_{\alpha}}}(\mean{\hat J_{\mathbf{n}}}_{\ket{\psi}}/{j'_{\alpha}})\\
&\ge& F_{j_{\alpha}}(\mean{\hat J_{\mathbf{n}}}_{\ket{\psi}}/{j_{\alpha}}),
\eean
where we have used Eq.~(\ref{eq:larger_j_is_better}) and $|c_{j_{\alpha}}|^2+|c_{{j'_{\alpha}}}|^2=1$.

Taking the maximum value of $j_\alpha$ 
({\it i.e.,} $j_\alpha=N_\alpha j$) we arrive at
\be \label{eq:Obs1proof1b}
\big(\Delta\hat J_{\mathbf{\perp}}^{(N_\alpha)} \big)^2 \ge 
N_\alpha j F_{N_\alpha j}\bigg(\frac{\mean{\hat J_{\mathbf{n}}^{(N_\alpha)}}}{N_\alpha j}\bigg) 
\ge N_\alpha j F_{k j}\bigg(\frac{\mean{\hat J_{\mathbf{n}}^{(N_\alpha)}}}{N_\alpha j}\bigg),
\ee
where the second inequality is due to (iii)
and $N_\alpha \le k$ for $k$-producible states.
Since the function $F_{j_\alpha}(X)$ 
is convex in $X$, we can now apply Jensen's inequality \cite{note2_obs1}
to the last term in Eq.~(\ref{eq:Obs1proof1b}).
We obtain that
\be \label{eq:Obs1proof1c}
\sum_{\alpha=1}^{M_N} \big(\Delta\hat J_{\mathbf{\perp}}^{(N_\alpha)} \big)^2 \ge
 N j F_{k j}\bigg(\frac{\mean{\hat J_{\mathbf{n}}^{(N)}}}{N j}\bigg),
\ee
where 
$\mean{\hat J_{\mathbf{n}}^{(N)}}=\sum_{\alpha} \mean{\hat J_{\mathbf{n}}^{(N_\alpha)}}$
for any $N$.
Finally, by combining Eqs.~(\ref{eq:Obs1proof1}) and (\ref{eq:Obs1proof1c}),
and using again Jensen's inequality \cite{note2_obs1}, we have
\be
	(\Delta\hat J_{\mathbf{\perp}})^2
	\ge \sum_N Q_{N} Nj\ F_{kj}\bigg(\frac{\mean{\hat J_{\mathbf{n}}^{(N)}}}{Nj}\bigg)
	\ge  \mean{\hat N}j F_{kj}\bigg(\frac{\mean{\hat J_{\mathbf{n}}}}{\mean{\hat N}j}\bigg),
	\nonumber
\ee
where $\mean{\hat N} = \sum_N Q_{N} N$ and $\mean{\hat J_{\mathbf{n}}} = \sum_N Q_{N} \mean{\hat J_{\mathbf{n}}^{(N)}}$.
This proves Obs. 1 [cf. Eq.~(\ref{eq:Obs1})].
\proofend

\subsection*{B. Example: symmetrization creates spin squeezing} 

Let us consider the state 
\be
\ket{\psi_\alpha}=\sqrt{\alpha}\ket{11}+\sqrt{1-\alpha}\ket{01}=(\sqrt{\alpha}\ket{1}+\sqrt{1-\alpha}\ket{0})\otimes\ket{1}
\ee
which is clearly separable. 
Therefore, it has a spin squeezing parameter $\xi^2\ge 1$
for any $\alpha$ and any combination of the directions $\mathbf{n}$ and $\mathbf{\perp}$,
{\it i.e.}, it is not spin squeezed. In particular, for the directions
$\mathbf{n} = \mathbf{z}$ and $\mathbf{\perp}=\mathbf{x}$ it is given by [cf. Eq.~(\ref{eq:SSP})]
\be\label{eq:xi2_a}
	\xi^2_\alpha = \frac{1}{\alpha^2}-\frac{2}{\alpha}+2\ge 1.
\ee

\begin{figure}[t!]
\begin{center}
\includegraphics[clip,scale=0.9]{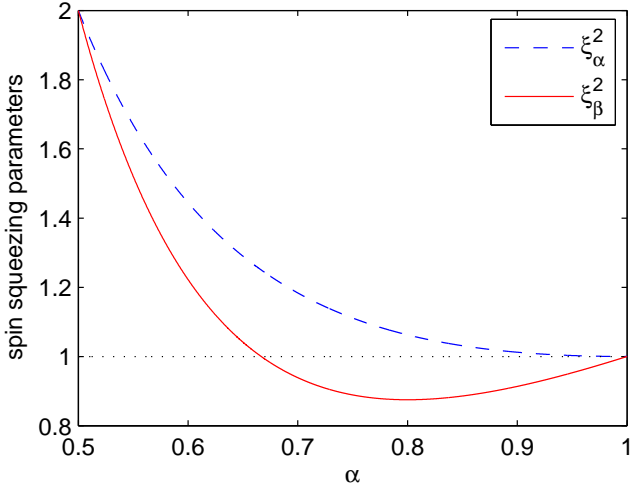}
\end{center}
\caption{Plot of the spin squeezing parameters $\xi_\alpha^2$ 
[dashed line, Eq.~(\ref{eq:xi2_a})] and $\xi_\beta^2$ 
[solid line, Eq.~(\ref{eq:xi2_b})]. 
While $\xi_\alpha^2$ 
never goes below the value 1 [dotted line], $\xi_\beta^2$ 
is below 1 for a large interval.
}  \label{fig:symmetrized_state}
\end{figure}

The corresponding state which could
be realized in the scenario of indistinguishable bosons is 
obtained by symmetrizing (and normalizing)
the state $\ket{\psi_\alpha}$, leading to 
\be \label{psiS}
\ket{\psi_\alpha^S}=\sqrt{\beta} \, \ket{11}+\sqrt{1-\beta} \, \frac{\ket{01}+\ket{10}}{\sqrt{2}},
\ee
where $\beta=\frac{2\alpha}{1+\alpha}$. 
For this state, the spin squeezing parameter 
for the same particular directions $\mathbf{n}=\mathbf{z}$ and $\mathbf{\perp}=\mathbf{x}$
is given by 
[cf. Eq.~(\ref{eq:SSP})]
\be\label{eq:xi2_b} 
	\xi_\beta^2=\frac{2}{\alpha^2}-\frac{5}{\alpha}+4.
\ee
This is smaller than the critical 
value 1 for a larger range of parameters.
The minimum is reached at $\xi_{\beta}^2=7/8$ for $\beta=4/5$,
which corresponds to $\alpha=2/3$.
Therefore, symmetrization does not preserve neither entanglement
nor spin squeezing.
In order to illustrate
the results, the two curves are plotted
in Fig.~\ref{fig:symmetrized_state}.

\subsection*{C. Proof of Observation 3}

Let us first introduce the formalism used in the proof.
Recall that we are considering here $N$ particles with $d_{\rm in}$ ($d_{\rm ex}$) 
internal (external)
degrees of freedom, labeled by $i=1,...,d_{\rm in}$ ($\gamma=1,...,d_{\rm ex}$).
Referring to Fig.~(\ref{fig:k_particle_entangled_states}), 
we can think of the external states as the energy levels of a spatial trap.
Each level $\gamma$ contains $N_\gamma$ particles in the state $\ket{\gamma}_{\rm ex}$.
These particles can be in different internal states $\ket{i}_{\rm in}$. 
In general, we have $N_{i,\gamma}$ particles in the state  
$\ket{i,\gamma}\equiv\ket{i}_{\rm in}\otimes\ket{\gamma}_{\rm ex}$
(to simplify the notation, we remove here the tensor product sign and pendices ``${\rm in}$'' and 
``${\rm ex}$'').
We also introduce a vector $\mathbf{N_{\rm ex}} = (N_1,N_2,...,N_{d_{\rm ex}})$
giving the occupation numbers of each external state
and a vector 
$\mathbf{N_\gamma} = (N_{1,\gamma} , N_{2,\gamma} ,..., N_{d_{\rm in},\gamma})$
with occupations of the internal states for a fixed external level $\gamma$.
Here, the relations $\sum_{\gamma=1}^{d_{\rm ex}} N_\gamma= N$
and $\sum_{i=1}^{d_{\rm in}} N_{i,\gamma} = N_\gamma$ hold.

Let us consider a specific example for 
$d_{\rm in}=d_{\rm ex}=2$ and $N=3$.
Choosing $N_{1,1}=1$, $N_{2,1}=0$, $N_{1,2}=2$ and $N_{2,2}=0$, 
we obtain $\mathbf{N_{\rm ex}}=(1,2)$,
$\mathbf{N}_1 = (1,0)$ and $\mathbf{N}_2 = (2,0)$.
The (non-symmetric) state is
\be\label{eq:AppCex1}
	\otimes_{\gamma=1}^2\otimes_{i=1}^2\ket{i,\gamma}^{\otimes N_{i,\gamma}}
  =\ket{1,1}\otimes\ket{1,2}\otimes\ket{1,2}. 
\ee
Finally, $\{ \mathbf{N_\gamma} \}_\gamma$ 
is the complete set of occupation numbers $\mathbf{N_\gamma}$ 
for all the $\gamma$ levels.
The corresponding symmetrized states with occupation numbers $\mathbf{N_\gamma}$ is
given by 
\be\label{eq:Dbasis}
	\ket{D_{\mathbf{N_{\rm ex}}}^{\{\mathbf{N_\gamma}\}}}\equiv
	\frac{1}{\sqrt{\cal N}}
	{\sum_\pi}\hat P_\pi\big[\otimes_{\gamma=1}^{d_{\rm ex}}\otimes_{i=1}^{d_{\rm in}}
	\ket{i,\gamma}^{\otimes N_{i,\gamma}}\big],
\ee
where $\hat P_\pi$ is a representation of 
the permutation $\pi$, and the sum runs through all {\it distinct} permutations,
the number of which is
${\cal N}\equiv{N\choose \{ \mathbf{N_\gamma} \}}\equiv\frac{N!}{\Pi_{i,\gamma} N_{i\gamma}!}$.
The states 
$\ket{D_{\mathbf{N_{\rm ex}}}^{\{\mathbf{N_\gamma}\}}}$ form a basis which is 
the analogous to a Fock state basis in second quantization.

The non-symmetric state from the 
example above [cf. Eq.~(\ref{eq:AppCex1})] 
becomes
\bean
	\ket{D_{\mathbf{N_{\rm ex}}}^{\{\mathbf{N_\gamma}\}}}&=&
	 \frac{1}{\sqrt{3}} 
	\big(
	\ket{1,1}\otimes\ket{1,2}\otimes\ket{1,2}+ \nonumber \\
	&& + \ket{1,2}\otimes\ket{1,1}\otimes\ket{1,2}+
	\ket{1,2}\otimes\ket{1,2}\otimes\ket{1,1} \nonumber
	\big).
	\nonumber
\eean

We will use the label $D$ in general for Fock states with a fixed 
occupation in internal and external levels in first quantization. In particular,
we employ symmetric states with $N_\gamma$ particles in the single
external level $\gamma$,
\be
\ket{D_{N_\gamma}^{\mathbf{N_\gamma}}}
\equiv\ket{I_{N_\gamma}^{\mathbf{N_\gamma}}}\otimes \ket{\gamma}^{\otimes N_\gamma},
\ee
where 
\be \label{eq:Obs2onegamma}
\ket{I_{N_\gamma}^{\mathbf{N_\gamma}}}\equiv
\frac{1}{\sqrt{{\cal N}_\gamma}}
{\sum_\pi} \hat P_\pi\Big[\otimes_i \ket{i}^{\otimes N_{i,\gamma}}_\gamma\Big],
\ee
and ${\cal N}_\gamma\equiv
{N_\gamma \choose \mathbf{N_\gamma}}\equiv\frac{N_\gamma!}{\Pi_{i} N_{i\gamma}!}$
is the number of distinct permutations $\pi$.
We attach the label $\gamma$ to $\ket{i}$
in order to keep track of the external level $\gamma$
the particle is in. This will be important below.

With these definitions, we can reformulate Observation 3 in technical terms.

\noindent
{\bf Observation 3}. 
{\it For any permutationally invariant
operator $\hat A_{\rm in}$ acting on the internal degree of freedom, and for a 
symmetric state 
$\ket{\Psi^{ \mathbf{N_{\rm ex}} }_{\rm S}}=
\sum_{\{ \mathbf{N_\gamma} \}}c_{\{ \mathbf{N_\gamma} \}}
\ket{D_{ \mathbf{N_{\rm ex}} }^{\{ \mathbf{N_\gamma} \}}}$ 
with a fixed occupation vector $\mathbf{N_{\rm ex}}$,}
\be
\label{eq:sym}
\bra{\Psi^{ \mathbf{N_{\rm ex}} }_{\rm S}}\hat A_{\rm in}\otimes\Eins_{\rm ex}\ket{\Psi^{\mathbf{N_{\rm ex}}}_{\rm S}}
=\bra{\Psi^{ \mathbf{N_{\rm ex}} }_{\rm in}}
\hat A_{\rm in}\ket{\Psi^{ \mathbf{N_{\rm ex}} }_{\rm in}}
\ee 
{\it holds, where
$\ket{\Psi^{ \mathbf{N_{\rm ex}} }_{\rm in}}=\sum_{\{\mathbf{N_\gamma}\}}
c_{\{\mathbf{N_\gamma}\}}
\big[\otimes_{\gamma}\ket{I_{N_\gamma}^{\mathbf{N_\gamma}}}\big]$, 
and $\ket{I_{N_\gamma}^{\mathbf{N_\gamma}}}$ is a symmetric
internal state as defined in Eq.~(\ref{eq:Obs2onegamma}).}

{\it Proof.} 
By inserting the definitions of 
$\ket{\Psi^{\mathbf{N_{\rm ex}}}_S}$ and $\ket{\Psi^{\mathbf{N_{\rm ex}}}_{\rm in}}$
it is easy to see that Eq.~(\ref{eq:sym}) holds if 
\be\label{eq:Obs2condition}
\bra{D_{{\mathbf{N_{\rm ex}}}}^{\{ \mathbf{N_\gamma} \}}} \hat A_{\rm in}\otimes\Eins_{\rm ex}
\ket{D_{{\mathbf{N_{\rm ex}}}}^{\{ \mathbf{N'_\gamma} \}}}
=\big[\otimes_\gamma\bra{I_{N_\gamma}^{\mathbf{N_\gamma}}}\big]\ \hat A_{\rm in}\ 
\big[\otimes_{\gamma'}\ket{I_{N'_{\gamma'}}^{\mathbf{N'_{\gamma'}}}}\big]
\ee
is true for all $\{ \mathbf{N_\gamma} \}$ and $\{ \mathbf{N'_{\gamma'}} \}$ 
with the same $\mathbf{N_{\rm ex}}$. 
We will show now that this is the case. We insert into the left
hand side of Eq.~(\ref{eq:Obs2condition}) the definition of the states 
$\ket{D_{{\mathbf{N_{\rm ex}}}}^{\{ \mathbf{N_\gamma} \}}}$
[cf. Eq.~(\ref{eq:Dbasis})], which leads to
\be \label{eqDD1}
\sum_{\pi,\pi'}
\Big[\frac{\hat P_\pi}{\sqrt{\cal N}} \otimes_{\gamma,i}\ket{i,\gamma}^{\otimes N_{i,\gamma}}\Big]^\dagger
\hat A_{\rm in}\otimes\Eins_{\rm ex}
\Big[\frac{\hat P_{\pi'}}{\sqrt{N'}}\otimes_{\gamma',i'}\ket{i',\gamma'}^{\otimes N_{i',\gamma'}'}\Big].
\ee
As before, we consider the sum of distinct permutations only.
Due to the identity $\Eins_{\rm ex}$ on the external states the terms
in the sum will vanish unless the $N_\gamma$ particles in level 
$\gamma$ are on the same positions in the permutations on both 
sides of $\hat A_{\rm in}\otimes \Eins_{\rm ex}$, since 
$\bra{i,\gamma}\hat A_{\rm in}\otimes\Eins_{\rm ex}\ket{i',\gamma'}=
\bra{i}\hat A_{\rm in}\ket{i'}\delta_{\gamma,\gamma'}$. 
Therefore, we can rewrite expression~(\ref{eqDD1}) as 
\bea \label{eq:AppC_sum1}
\sum_{\pi,\tilde\pi,\tilde\pi'}
\Big[\frac{\hat P_{\pi}\hat P_{\tilde\pi}}{\sqrt{\cal N}}\otimes_{\gamma,i}\ket{i}^{\otimes N_{i,\gamma}}_\gamma\Big]^\dagger
\hat A_{\rm in}\Big[\frac{\hat P_{\pi}\hat P_{\tilde\pi'}}{\sqrt{\cal N'}}
\otimes_{\gamma',i'}\ket{i'}^{\otimes N_{i',\gamma'}'}_\gamma\Big].\nonumber\\
\eea
Here the permutations $\tilde\pi$ and $\tilde\pi'$ permute particles stemming
from the same external state $\gamma$, and 
$\pi$ permutes particles with a different $\gamma$.
Note that for simplicity we use the same 
operators $\hat P_{\tilde\pi}$ to represent
a permutation $\tilde\pi$ of the $N$ particles 
even though now the state 
space of each particle is reduced to 
the internal states. In order to clarify the notation we employed,
we note that for the example considered in Eq.~(\ref{eq:AppCex1}),
the reduced state would be
\be
\otimes_{\gamma,i}\ket{i}_\gamma^{\otimes N_{i,\gamma}}=
	\ket{1}_1
	\otimes 
	\ket{1}_2^{\otimes 2}
	=
	\ket{1}_1
	\otimes 
	\ket{1}_2
  \otimes
  \ket{1}_2.\nonumber
\nonumber
\ee
Since $\hat A_{\rm in}$ is permutationally invariant, we have that 
$\hat P_\pi^\dagger\hat A_{\rm in}\otimes
\Eins_{\rm ex}\hat P_\pi=\hat A_{\rm in}\otimes
\Eins_{\rm ex}$. Hence in the sum over $\pi$ each term contributes 
equally, and the sum  can be 
replaced by the number of distinct permutations 
${\cal N}_{\rm ex}\equiv{N\choose \mathbf{ N_{\rm ex}} }\equiv\frac{N!}{\Pi_{\gamma} N_{\gamma}!}$
in expression~(\ref{eq:AppC_sum1}).

We observe that 
because the permutations $\tilde\pi$ only permute particles with the same $\gamma$,
one can rewrite  
$\sum_{\tilde\pi}\hat P_{\tilde\pi}=\Pi_\gamma \big(\sum_{\pi_\gamma}\hat P_{\pi_\gamma}\big)$
of permutations $\pi_\gamma$ which permute particles in the level $\gamma$.
This leads to
\bean
\sum_{\tilde\pi}\hat P_{\tilde\pi}\big[\otimes_{\gamma,i}\ket{i}^{\otimes N_{i,\gamma}}_\gamma\big]
&=&\otimes_{\gamma}\sum_{\pi_\gamma}\hat P_{\pi_\gamma}\big[\otimes_{i}\ket{i}^{\otimes N_{i,\gamma}}_\gamma\big]\\
&=&\sqrt{\Pi_\gamma{\cal N}_\gamma} \big[\otimes_\gamma\ket{I_{N_\gamma}^{\mathbf{N_\gamma}}}\big],
\eean
cf. Eq.~(\ref{eq:Obs2onegamma}). We arrive at
\be
\sqrt{\frac{{\cal N}_{\rm ex}^2\ \Pi_\gamma{\cal N}_\gamma\ \Pi_{\gamma'}{\cal N}'_{\gamma'}}
{{\cal N}{\cal N}'}}
\big[\otimes_\gamma\bra{I_{N_\gamma}^{\mathbf{N_\gamma}}}\big]\ \hat A_{\rm in}\ 
\big[\otimes_{\gamma'}\ket{I_{N_{\gamma'}}^{\mathbf{N'_{\gamma'}}}}\big].\nonumber
\ee
One can directly check that the prefactor is equal to $1$. Therefore, condition
Eq.~(\ref{eq:Obs2condition}) is fulfilled. \proofend

\subsection*{D. Dilute cloud argument}

\begin{figure}[t!]
\begin{center}
\includegraphics[clip,scale=0.6]{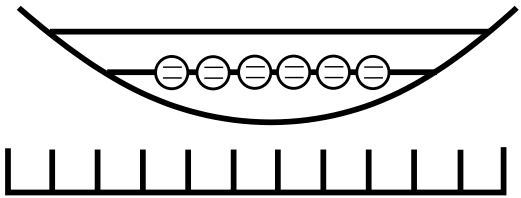}
\end{center}
\caption{Initially all particles are in the lowest 
energy level of the trap. 
The cloud is then released.
If it is diluted enough along the horizontal direction,
it is likely that at most one particle falls into each of the 
boxes, which represent the single detectors.
}  \label{fig:dilute_cloud}
\end{figure}

One may think that, after preparing the Bose-Einstein condensate atoms
in the ground state of a confining trap, it is possible to apply the spin squeezing
inequalities (\ref{SSI_multi}) and (\ref{eq:Obs1})
by simply
releasing the trap, letting the cloud expand and 
fall onto a grid of small detectors capable of measuring the internal 
state of a single atom. 
If the cloud is dilute enough, 
it is very likely that at most a single atom enters each detector, 
thus making the atoms distinguishable. 
This would make it possible to apply the S{\o}rensen-M{\o}lmer 
bounds. The situation is illustrated in Fig.~\ref{fig:dilute_cloud}.
We show here that this argument, which is often encountered in discussions, does 
resolve the problem.

Let us assume that, before releasing the atoms from the trap,
their state is of the form 
\be \label{Arg1.initialstate}
	\ket{\Psi}=\ket{\psi_{\rm S}}_{\rm in}\otimes\ket{0}_{\rm ex}^{\otimes N},
\ee
where $\ket{\psi_{\rm S}}_{\rm in}$ is a symmetric total internal state
and each atom is in the  ground state of the trap $\ket{0}_{\rm ex}$. 
Here ``ex'' (``in'') indicates the external 
(internal) degree of freedom as in Section \ref{sec:Operational_Interpretation}. 
We assume here that all atoms share the same spatial wave function, which 
can thus be factorized.
If interactions can be neglected during free fall, then only the spatial
state of each atom changes, leaving the 
internal state symmetric.
By waiting long enough, the single-particle spacial wave function becomes so 
spread that the probability to detect two atoms at the same spatial detector is negligible.

Let us assume for simplicity that the atoms are trapped and detected 
state-insensitively first, such that at most one atom is detected in each 
site. A problem is that in each shot, different sites will be occupied. 
This might still be considered as a minor problem. In a
one-dimensional trap,
for instance, it could be resolved by identifying particle ``1''
with the leftmost trap, particle ``2'' with the particle 
right from particle ``1'', and so on. Alternatively, one could
postselect on events where always the same $N$ sites are occupied.

In general, the position measurement makes 
the state effectively distinguishable.
Let us illustrate the situation with an example for $N=2$ particles, labeled as $1$ and $2$,
in two different sites labeled by $a$ and $b$. 
As in Section \ref{sec:Operational_Interpretation},
we consider a general pure symmetric state
\be\label{eq:ex_N2}
	\ket{\psi}=\frac{1}{\sqrt{2}}\big(\ket{\psi_{12}}_{\rm in}\otimes\ket{a_1 b_2}_{\rm ex}+\ket{\psi_{21}}_{\rm in}\otimes\ket{b_1 a_2}_{\rm ex}\big),
\ee
with the same definitions as in Eq.~(\ref{eq:general_state}).
An operator
$\hat M_a$ acting on the internal state of the particle in site $a$ can be written as
\bean
	\hat M_a&=&(\hat A_1\otimes \Eins_2)_{\rm in}\otimes(\hat m^a_1 \otimes\Eins_2)_{\rm ex}\\
	        &+&(\Eins_1\otimes\hat A_2)_{\rm in}\otimes(\Eins_1\otimes \hat m^a_2)_{\rm ex},
\eean
where $\hat m^a\ket{a}=\ket{a}$ and $\hat m^a\ket{b}=0$ since we measure locally
at site $a$. $\hat M_a$ 
has to be permutationally invariant with respect to the interchange of the particle
labels since the particles are indistinguishable \cite{PeresBook}. 
The expectation value 
with respect to the state (\ref{eq:ex_N2}) is
\be
	\bra{\psi}\hat M_a\ket{\psi}=\frac{1}{2}[\bra{\psi_{12}}\hat A_1\otimes\Eins_2\ket{\psi_{12}}
	             +\bra{\psi_{21}}\Eins_1\otimes\hat A_2\ket{\psi_{21}}],
\ee
where the two terms are equal since 
\bean
\bra{\psi_{21}}\Eins_1\otimes\hat A_2\ket{\psi_{21}} &=& 
\bra{\psi_{12}} \hat{P}_{\rm in}^\dag (\Eins_1\otimes\hat A_2) \hat{P}_{\rm in} \ket{\psi_{12}} \\
&=& \bra{\psi_{12}}\hat A_1\otimes\Eins_2\ket{\psi_{12}}.
\eean
We dropped the label ``in'' of $\ket{\psi_{12}}_{\rm in}$
for simplicity.
An analogous result is obtained when considering an 
operator acting on the internal state of the particle on site $b$.
Since only such operators are measured in the usual
scenario, we can identify particle 1 with site $a$ and particle 2 with 
site $b$, and $\ket{\psi_{12}}$ is sufficient to describe 
the state of the two particles. This is a state of two distinguishable
particles.

However, since the measurement acts only on the external degrees of freedom,
the product structure between the internal and the external degrees of freedom
in Eq.~(\ref{Arg1.initialstate}) is preserved.
It is then evident that the internal state 
remains fully symmetric even after the position 
measurement, since this affects the external state only. 
Hence making the state distinguishable effectively
after the state transformation in the measurement does not make 
it possible to leave the restricted class of symmetric states:
in the interferometric situation we considered above, when 
only the sites $a$ and $b$ are occupied, then we arrive at the 
state of Eq.~(\ref{eq:ex_N2}), but with a symmetric internal state 
$\ket{\psi_{12}}$.
Therefore, the effective state of the distinguishable 
particles is symmetric.  
This example, which can be directly generalized 
to $N$ particles, shows that simply making the cloud dilute does 
not make it possible to apply the spin squeezing bounds discussed in this 
paper.




\begin{thebibliography}{99}
\bibitem{noteRamsey} 
E.g., for the Ramsey-type interferometer, we have
$\mathbf{m} = \mathbf{y}$, 
$\mathbf{n} = \mathbf{x}$ 
and ${\mathbf{\perp}} = \mathbf{z}$.
\bibitem{notaSpinSq} 
A different generalization of Eq.~(\ref{eq:SSP}), 
using normalized spin operators $\hat J_{\mathbf{n}}/ \hat N$ and 
$\hat J_{\mathbf{\perp}}/ \hat N$, has been proposed in Ref.~\cite{HeArXiv11}.
\bibitem{noteSM} 
The functions $F_j(X)$ are essentially constructed by calculating 
the mean value $\mean{\hat j_{\mathbf{n}}}$ and the
variance $(\Delta \hat j_{\mathbf{\perp}})^2$
for the eigenstates of the operator $\mu \hat j_{\mathbf{n}} + \hat j_{\mathbf{\perp}}^2$, 
where $\mu$ is a Lagrange multiplier.
For details see Ref.~\cite{SoerensenPRL01}.
\bibitem{nota_X}
Note that $X\ge 0$ is assumed without loss of generality.
\bibitem{nota_modes}
Other authors favour the use of a mode-picture 
\cite{Benatti}, even though there is 
no direct connection between entanglement and the 
precision in this case except for special situations
\cite{HeArXiv11}.
\bibitem{nota_indist}
There is a large body of literature on general
entanglement of indistinguishable particles.
For a recent review, see \cite{AmicoRMP08}.
\bibitem{nota_optim}
We minimize the variance of $\hat J_x$, which is a concave function, 
over the set of $k$-producible states ${\cal S}_k$ with some mean
value $\mean{\hat J_z}$, which is a convex set.
For such a problem, the minimum
is reached on the pure states \cite{BoydBook}.
\end{thebibliography}
\end{document}